\documentclass[10pt]{iopart}

\expandafter\let\csname equation*\endcsname\relax
\expandafter\let\csname endequation*\endcsname\relax
\usepackage{amsmath}
\usepackage{amssymb}
\usepackage{color}
\usepackage{graphicx}
\usepackage{hyperref}
\usepackage{iopams}
\usepackage{ulem}

\newcommand{\orcid}[1]{\href{https://orcid.org/#1}
{\includegraphics[width=7pt]{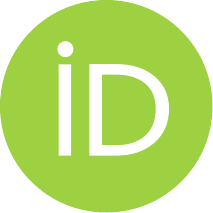}}}

\hypersetup{
  colorlinks=true,linkcolor=blue,citecolor=blue,
  filecolor=blue,urlcolor=blue,breaklinks=true
}

\begin{document}
\title[]{%
  Entanglement dynamics of two optical modes coupled through a
  dissipative movable mirror in an optomechanical system
}

\author{Bruno P. Schnepper\orcid{0000-0003-1159-3166}}
\ead{b.piveta@ufabc.edu.br}
\address{
  Centro de Ci\^encias Naturais e Humanas,
  Universidade Federal do ABC,
  09210-580 Santo André, São Paulo, Brazil
}

\author{Danilo Cius\orcid{0000-0002-4177-1237}}
\ead{danilocius@gmail.com}
\address{
  Departamento de F\'{i}sica Matem\'{a}tica,
  Instituto de F\'{i}sica da Universidade de S\~{a}o Paulo
  05508-090 São Paulo, São Paulo, Brazil
}

\author{Fabiano M. Andrade\orcid{0000-0001-5383-6168}}
\ead{fmandrade@uepg.br}
\address{
  Programa de P\'os-Gradua\c{c}\~{a}o Ci\^{e}ncias/F\'{i}sica,
  Universidade Estadual de Ponta Grossa,
  84030-900 Ponta Grossa, Paran\'a, Brazil
}
\address{
  Departamento de Matem\'{a}tica e Estat\'{i}stica,
  Universidade Estadual de Ponta Grossa,
  84030-900 Ponta Grossa, Paran\'a, Brazil
}

\begin{abstract}
Nonclassical states are an important class of states in quantum mechanics, particularly for applications in quantum information theory. Optomechanical systems are invaluable platforms for exploring and harnessing these states. In this study, we focus on a mirror-in-the-middle optomechanical system. In the absence of losses, a separable state, composed of the product of coherent states, evolves into an entangled state. Furthermore, we demonstrate that generating a two-mode Schrödinger-cat state depends on the optomechanical coupling. Additionally, when the optical modes are uncoupled from the mechanical mode, we find no entanglement for certain nonzero optomechanical coupling intensities. We exactly solve the Gorini-Kossalokowinki-Sudarshan-Lindblad master equation,  highlighting the direct influence of the reservoir on the dynamics when mechanical losses are considered. Then, we discuss vacuum one-photon superposition states to obtain exact entanglement dynamics using concurrence as a quantifier. Our results show that mechanical losses in the mirror attenuate the overall entanglement of the system.\\

\noindent DOI:
\href{https://doi.org/10.1088/1751-8121/ad92d0}
{10.1088/1751-8121/ad92d0}
\end{abstract}

\vspace{2pc}
\vspace{2pc}

\submitto{\JPA}

\section{Introduction}
The entanglement phenomenon has been discussed since the introduction of
quantum mechanics \cite{schrodinger1935}.
This phenomenon occurs when the quantum states of two or more particles
are correlated such that the state of one particle cannot be
described independently of the state of the others, even when the
particles are spatially separated.
It is traditionally understood to be a property of non-separability
\cite{horodecki2009}.
It implies that information about the state of one particle is somehow
connected to the state of another, which challenges the classical notion
of locality \cite{einstein1935}.
In the sense of Bell nonlocality \cite{bell1964}, not all entangled
state violates Bell's inequality. However, each state that violates it
is entangled.
Therefore, entanglement is a necessary condition for violating Bell's
inequality \cite{werner1989}.
These concepts together highlight the nonclassical nature of quantum
mechanics and have profound implications for our understanding of
reality.

The importance of quantum entanglement  extends beyond the foundations
of quantum mechanics and is recognized as a valuable resource for
quantum cryptography \cite{ekert1992}, quantum teleportation
\cite{bennett1993}, quantum metrology \cite{giovannetti2011}, and
quantum control \cite{guha2023}.
Recently, a proposal for an entanglement-based protocol to test
short-distance quantum physics, such as the magnetically induced
dipole-dipole interaction and the Casimir-Polder potential between two
nano-crystals in a nonrelativistic regime was demonstrated
\cite{marshman2024}.
It facilitates secure communication \cite{nadlinger2022} and offers a
promising avenue to build quantum computers \cite{raussendorf2001} that
can solve certain mathematical problems more efficiently
\cite{arute2019}.
Consequently, a surge in interest is evident concerning theoretical
advancements and the creation of experimental instruments for generating
entangled states.
It extends across diverse interfaces and platforms, from microscopic
systems to mesoscopic devices.
Examples include atomic/molecular systems \cite{raimond2001},
superconductor circuits \cite{you2011}, and photonic \cite{pan2012}.
In particular, optomechanical systems are capable of achieving
entanglement in massive objects \cite{aspelmeyer2014}, which makes this
kind of system, a resourceful platform to explore nonclassicalities and
experiments for probing the gravitational effects of quantum mechanical
matter \cite{carney2019,biswas2023}.

In optomechanical systems,  light interacts with a mechanical element,
enabling indirect manipulation of the mechanical state.
When the mechanical element acts as one of the mirrors of the cavity and
moves along an axis, its position determines the resonant frequency of
the cavity mode \cite{bowen2015}.
The photons momenta cause slight displacements of the mechanical
element, altering the cavity length and optical frequency.
This change in radiation pressure signifies a nonlinear interaction
between optical modes and mechanical displacement.
The nonlinear interaction is crucial for generating nonclassical effects
such as the generation of Schrödinger-cat state and optical squeezing
\cite{sougato1997}, which are crucial for the detection of gravitational
waves preceding a binary black hole coalescence, as those detected for
the first time in 2015, by the LIGO-Virgo collaboration
\cite{abbott2016},  and various quantum technologies
\cite{barzanjeh2022}.

In pursuit of the most realistic scenarios, considering open quantum
dynamics is pivotal in both experimental and theoretical explorations of
cavity optomechanical systems.
Addressing situations where the system is not isolated from its
environment involves employing frameworks such as quantum Langevin or
master equations \cite{gardiner2004}.
For instance, a perturbative solution to the master equation for
nonlinear optomechanical systems with optical loss was provided in
\cite{mancini1997}, while a recent elegant solution integrating a Lie
algebra approach with a vectorized representation of the Lindblad
equation was presented in \cite{qvarfort2019}.
The investigation of mechanical loss has been a topic of interest for
decades, with treatments ranging from master equation formulations
\cite{sougato1997} to considerations of Brownian motion
\cite{bassi2005}.
Moreover, an approach that accounts for optical and mechanical losses
within a damping-basis framework was presented in \cite{torres2019}.

Here, we consider the dynamics in an optomechanical system comprising two
optical cavities coupled to a mechanical oscillator.
This scheme called the mirror-in-the-middle configuration, is
represented in Fig. \ref{fig:scheme}.
In this configuration, the optical modes do not interact directly with
each other; the movable mirror indirectly mediates their interaction,
resulting in their entanglement.
We then focus specifically on scenarios where decoherence originating
from the damping of mechanical motion dominates over other sources, such
as photon leakage, which we consider negligible.
We derive an analytical expression for the time evolution of the density
operator in the Schrödinger picture, adopting the same ansatz solution
as used in Ref. \cite{sougato1997} and proceed to solve exactly the
associated differential equations.
Our findings represent an improvement on the reference above
due to methodological differences.
While the authors employed a technique alternating between unitary and
nonunitary evolution for brief intervals to solve the master equation,
our approach directly addresses the differential equations.
Consequently, we demonstrate that while the damping term remains
identical across both solutions, their treatment neglects the effect of
the reservoir on the coherent term.

This work is organized as follows.
In Sec. \ref{sec:unitary}, we analyze the mechanical motion that induces
entanglement between the optical and mirror states by preparing coherent
states in all partitions, with linear entropy applying as a quantifier
of entanglement.
The linear entropy oscillates between null and maximum positive values,
denoting the separability or entanglement of optical fields and mirror
states, respectively.
The mirror state becomes decoupled from the state of optical fields at
certain times.
At these times, we verify the generation of two-mode Schrödinger-cat
states and evaluate the entanglement of the state of optical fields by
calculating the linear entropy as a function of optomechanical coupling,
demonstrating the existence of nonnull optomechanical coupling values
that cause the separability of optical modes.
In Sec. \ref{sec:nonunitary}, we introduce mechanical loss and solve
analytically the master equation as mentioned before.
Then, we apply our exact solution of the master equation, considering
mechanical loss for analyzing the case when the fields are initially
prepared in vacuum one-photon superposition states \cite{lombardi2002}.
This preparation is interesting because the system dynamics become
restricted to a two-dimensional space spanned by the vacuum and the
one-photon states.
In this case, we can evaluate the concurrence to quantify entanglement
even when the composite system is in a mixed state, as presented in
\cite{brandao2020} for the unitary case.
Then, we compare the concurrence calculated from our exact density matrix
with the approximation obtained in Ref. \cite{sougato1997}.
The main conclusions and developments are presented in
Sec. \ref{sec:conclusion}.

\section{Unitary Dynamics}
\label{sec:unitary}

\begin{figure}[t]
  \centering
  \includegraphics[width=0.5\linewidth]{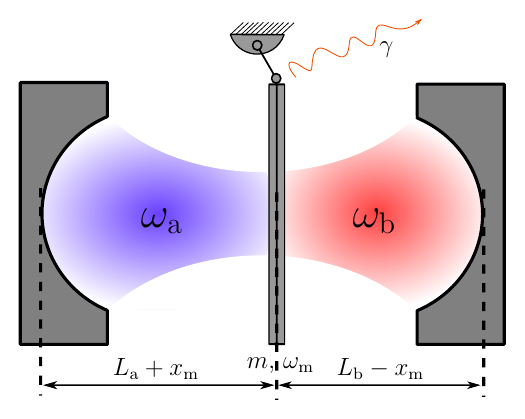}
  \caption{
    A schematic representation of optomechanical setup where the optical
    modes ($a$, $a^{\dagger}$) and ($b$, $b^{\dagger}$) are coupled to
    the mechanical mirror position $x_{\text{m}}$ via the interaction
    term $-(g_{\text{a}}a^{\dagger}a - g_{\text{b}}b^{\dagger}b)x_{\text{m}}$.
    Friction and imperfections cause loss of phonons from the mirror at a
    rate $\gamma_{\text{diss}}$, which we represent as a rescaled number
    concerning the mechanical frequency $\omega_{\text{m}}$ as $\gamma=\gamma_{\text{diss}}/\omega_{\text{m}}$.
  }
  \label{fig:scheme}
\end{figure}

The system configuration consists of two optical cavities with different
lengths, $L_{\text{a}}$ and $L_{\text{b}}$, each containing modes of
different frequencies, $\omega_{\text{a}}$ and $\omega_{\text{b}}$, and
a movable mirror with a mass $m$ subject to a harmonic potential of
frequency $\omega_{\text{m}}$.
The optical modes interact indirectly through the dispersive coupling
mediated by the mechanical mode.
A scheme of this configuration is illustrated in Fig. \ref{fig:scheme}.

Considering the coupling between the cavity field and the first power of
the mechanical displacement,  the Hamiltonian operator, which represents
the mirror-in-the-middle configuration is given by
\begin{align}
  \frac{H}{\hbar} = {}
  &
    \omega_{\text{a}} a^{\dagger} a
    + \omega_{\text{b}} b^{\dagger} b + \omega_{\text{m}} c^{\dagger}c
    - g_{\text{a}} a^{\dagger} a(c^{\dagger} + c)
    + g_{\text{b}} b^\dagger b(c^{\dagger} + c),
\end{align}
where $g_{\text{a},\text{b}} = \omega_{\text{a},\text{b}}
x_{\text{ZPF}}/L_{\text{a},\text{b}}$ are the optomechanical coupling
intensities, with $x_{\text{ZPF}} = \sqrt{\hbar/2m\omega_\text{m}}$
being the zero-point fluctuation of the mirror.
The operators $a$, $b$, and $c$  ($a^\dagger$, $b^\dagger$, and
$c^\dagger$) are the usual bosonic annihilation  (creation) operators
relative to each optical and mechanical mode, respectively.
In the absence of any dissipation ($\gamma=0$), the time-dependent
Schrödinger equation governs the time-evolution of the system,
$i\hbar \partial_{t} \vert \psi(t) \rangle = H  \vert \psi(t) \rangle$,
whose formal solution is represented by
$\vert \psi(t) \rangle = U(t) \vert \psi(0) \rangle$.
The initial state of the system $\vert \psi(0) \rangle$ evolves
deterministically to the state $\vert \psi(t) \rangle$ through the
time-evolution operator $U(t)$.
In this case, the time-evolution operator assumes the form
\begin{align}
  \label{eq:teo}
  U(t) =  {}
  &
    e^{-it(r_{\text{a}}a^{\dagger}a+r_{\text{b}}b^{\dagger}b) }
    e^{i(t-\sin t)(k_{\text{a}}a^{\dagger}a-k_{\text{b}}b^{\dagger}b)^{2}}
    e^{(k_{\text{a}}a^{\dagger}a-k_{\text{b}}b^{\dagger}b)[\eta(t)c^{\dagger}
    -\eta^{\ast}(t)c]}e^{-itc^{\dagger}c},
\end{align}
where we define the dimensionless coupling parameter
$k_{\text{a},\text{b}} = g_{\text{a},\text{b}}/\omega_{\text{m}}$, the
scaled time  $\omega_{\text{m}} t\rightarrow t$ and the parameters
$r_{\text{a},\text{b}}= \omega_{\text{a},\text{b}}/\omega_{\text{m}}$
and the time-dependent function $\eta(t) = 1 - e^{-it}$.
We note that an optically driven displacement operator appears to be
acting on the mechanical mode state in the time-evolution operator,
besides a  Kerr-like term between both optical modes.
The coupling between the optical modes and the mirror is proportional to
$k_{\text{a},\text{b}}$.
In contrast, the two optical modes indirectly interact by a second-order
term proportional to $k_{\text{a},\text{b}}^{2}$.
Remarkably, it is expected that nonclassical features spring during the
system evolution.
It is worth noting that this scenario is distinct from the production of
nonclassical states in a Kerr medium, which involves direct interaction
between two light modes \cite{chefles1996}.
Here, nonclassical correlations would appear as both optical modes
independently interact with the same movable mirror, hence indicating
the nonclassical nature of the mechanical mode \cite{marletto2017}.
Furthermore, if an initially separable state leads to the birth of
optical entanglement, there will inevitably be a subsequent death of
that entanglement.
For instance, starting from a completely separable state,
\begin{equation*}
\vert\psi(0)\rangle=\vert\psi_{\text{A}}\rangle
\otimes\vert\psi_{\text{B}}\rangle\otimes\vert\psi_{\text{C}}\rangle,
\end{equation*}
at times
$t = \tau_q \equiv 2q\pi$ ($q\in\mathbb{N}$), the function $\eta(t)$
vanishes for all $q$,  $\eta(\tau_{q})=0$, and therefore the state at
this time is given by
\begin{equation*}
\vert\psi(\tau_{q})\rangle =
e^{i\tau_{q}(k_{a}a^{\dagger}a-k_{b}b^{\dagger}b)^{2}}
[e^{-ir_{a}\tau_{q}a^{\dagger}a}\vert\psi_{\text{A}}\rangle
\otimes
e^{-ir_{b}\tau_{q}b^{\dagger}b}\vert\psi_{\text{B}}\rangle]
\otimes
e^{-i\tau_{q} c^{\dagger}c}\vert\psi_{\text{C}}\rangle,
\end{equation*}
where the optical modes A and B may
become entangled depending on the optomechanical coupling intensities $k_{\text{a},\text{b}}^{2}$.
In contrast, mode C is disentangled from AB.

\subsection{Coherent States}
Let us consider the case in which the system is initially prepared in a
separable state composed of the product of coherent states
\begin{equation}
    \vert\psi(0)\rangle = \vert\alpha\rangle \otimes
    \vert\beta\rangle \otimes \vert\phi\rangle.
\end{equation}
A coherent state can be defined as being a displacement of the vacuum
state in the phase space.
Mathematically, it is expressed as
$\vert\alpha\rangle = \hat{D}(\alpha)\vert 0 \rangle$, with
$\hat{D}(\alpha) = e^{\alpha a^{\dagger} - \alpha^{\ast} a}$ being the
displacement operator.
Using Eq. \eqref{eq:teo}, this state evolves in time as
\begin{align}
  \label{psit_coherent}
  \vert\psi(t)\rangle
  =
  &
    \sum_{m,n=0}^{\infty}c_{mn}(t)\vert m\rangle\otimes
    \vert n\rangle\otimes\vert\phi_{mn}(t)\rangle,
\end{align}
where
\begin{align}
  \label{c_mn}
  c_{mn}(t)
  =e^{-(|\alpha|^{2}+|\beta|^{2})/2}\frac{[\alpha(t)]^{m}}{\sqrt{m!}}
  \frac{[\beta(t)]^{n}}{\sqrt{n!}} e^{i\kappa_{mn}^{2}(t-\sin t)},
\end{align}
with
$\kappa_{mn}=k_{\text{a}}m-k_{\text{b}}n$,
$\alpha(t)=\alpha e^{-ir_{\text{a}}t}e^{ik_{\text{a}}\Im[\phi\eta(t)]}$,
$\beta(t)=\beta e^{ir_{\text{b}}t}e^{-ik_{\text{b}}\Im[\phi\eta(t)]}$,
and $\phi_{mn}(t)=\phi e^{-it}+\kappa_{mn}\eta(t)$.
It is clear from Eq. \eqref{psit_coherent} that the states of optical
and mechanical modes become correlated due to the coupling
$\kappa_{mn}$.
However, at the instants of time $t=\tau_{q}$ defined above, notably,
the optical modes become uncoupled from the mirror because the
displacement term responsible for coupling the optical and mechanical
modes vanishes since $\eta(\tau_{q})=0$.
In this case, we have
\begin{align}
  \label{eq:psitauq}
  \vert\psi(\tau_{q})\rangle
  = \vert \chi(\tau_{q})\rangle_{\text{AB}}\otimes\vert\phi\rangle_{\text{C}},
\end{align}
where the composite state of optical modes is
\begin{equation}
  \label{ABstate_q}
  \vert \chi (\tau_{q})\rangle_{\text{AB}}
  =\sum_{m,n=0}^{\infty}c_{mn}(\tau_{q}) \vert
  m\rangle\otimes
  \vert n\rangle.
\end{equation}
This state exhibits a nonclassical feature characterized by the
multi-component Schrödinger-cat state in two modes, which emerges
depending on the coupling intensity.
A Schrödinger-cat state in two modes is entangled.
To elucidate this fact, we assume the coupling constants as being
$k_{\text{a}}\approx k_{\text{b}}=\kappa$.
Thus, for example, by setting $\kappa=1/2$, the state is given in
Eq. \eqref{ABstate_q} can be written as
\begin{align}
  \vert \chi(\tau_{q})\rangle_{\text{AB}} = {}
  &
    \frac{1+e^{i\tau_{q}/4}}{2}
    |\alpha(\tau_{q})\rangle \otimes|\beta(\tau_{q})\rangle
    +
    \frac{1-e^{i\tau_{q}/4}}{2}
    |-\alpha(\tau_{q})\rangle
    \otimes
    |-\beta(\tau_{q})\rangle,
\end{align}
which corresponds to a two-component Schrödinger-cat state in two modes
for odd $q$ values.
Moreover, three- and four-component Schrödinger-cat states are
generated, respectively, for $\kappa=1/\sqrt{6}$ and
$\kappa=1/(2\sqrt{2})$, in agreement with the results reported in
Ref. \cite{sougato1997}.

The quantum entanglement between optical and mechanical modes is a
nonclassical property worth analyzing.
Indeed, as we have a tripartite system, it is possible to analyze the
entanglement between the bipartition composed of the optical and
mechanical modes are labeled AB and C, respectively.
In that case, the pure density matrix
$\rho_{\text{AB,C}}(t)=\vert\psi(t)\rangle\langle\psi(t)\vert$
represents the state of the system described by Eq. \eqref{psit_coherent}.
We may apply the von Neumann entropy \cite{petz2001}, a quantifier for
entanglement in pure bipartite states, to evaluate their entanglement.
It is defined as
\begin{equation}
  \mathcal{S}(\rho_{i})=-\Tr(\rho_{i}\ln\rho_{i}),
\end{equation}
for the reduced state $\rho_{i}$ $(i=\text{AB, C})$.
Specifically, it yields a null entropy value for separable states,
indicating their lack of entanglement.
Conversely, for entangled states, the von Neumann entropy returns a
positive value, denoting the presence of nonclassical correlations
within the system.

Nevertheless, evaluating the von Neumann entropy is difficult due to the
logarithm function.
Hence, instead of applying the von Neumann entropy for this purpose, we
employ the linear entropy as a quantifier of entanglement.
The linear entropy is given by
\begin{equation}
  \mathcal{S}_{\text{L}}(\rho_{i})=1-\Tr\,\rho_{i}^{2},
\end{equation}
which yields $\mathcal{S}_{\text{L}}(\rho_{i})=0$ for separable states
and $\mathcal{S}_{\text{L}}(\rho_{i})>0$ for entangled states.
The advantage of applying linear entropy to quantify entanglement is
the achievement of simple analytical expression in terms of the purity
of the reduced state $\rho_{i}$, which is given by $\Tr\,\rho_{i}^{2}$.
Purity belongs to the interval $[0,1]$, which equals $1$ for a pure
state and less than the unity for a mixed state.
Moreover, in the absence of losses, the linear entropy is symmetric to
the partitions, which means that
$\mathcal{S}_{\text{L}}(\rho_{\text{AB}})
=\mathcal{S}_{\text{L}}(\rho_{\text{C}})$.
In this manner, using the state in Eq.
\eqref{psit_coherent}, we obtain the analytical expression
\begin{equation}
  \label{SL_CS}
  \mathcal{S}_{\text{L}}(\rho_{i}(t)) = 1
  -\sum_{k,l,m,n=0}^{\infty}|c_{kl}|^2|c_{mn}|^2
  e^{-(\kappa_{kl}-\kappa_{mn})^{2}|\eta(t)|^2},
\end{equation}
with $|c_{kl}(t)|^2=|c_{kl}(0)|^2=|c_{kl}|^2$ given by Eq. \eqref{c_mn}.
\begin{figure}[t]
  \centering
  \includegraphics[width=0.75\linewidth]{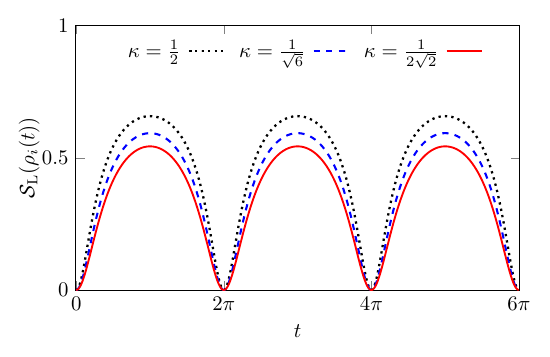}
  \caption{
    Linear entropy of state of the $i$-th partition
    ($i=\text{AB},\text{C}$) as a function of the dimensionless times
    $t$ considering $\alpha=\beta=1$.
    In this case, the linear entropy quantifies the degree of
    entanglement between the two optical and mechanical modes for
    different coupling intensities $k_{\text{a}}\approx
    k_{\text{b}}=\kappa$, where $\kappa=1/2$ (black dotted line),
    $\kappa=1/\sqrt{6}$ (blue dashed line), and  $\kappa=1/(2\sqrt{2})$
    (red solid line).
    The linear entropy vanishes at time $t=\tau_{q}=2q\pi$
    ($q\in\mathbb{N}$), which confirms that the state of the system is
    given by Eq. \eqref{eq:psitauq}.
  }
   \label{fig:SL_CS}
\end{figure}
In Fig. \ref{fig:SL_CS}, the behavior of linear entropy is plotted for
different values of coupling intensities
$k_{\text{a}}\approx k_{\text{b}}=\kappa$, considering $\alpha=\beta=1$.
In this case, the linear entropy shows the birth and death of quantum
entanglement between the two optical and mechanical modes for different
coupling intensities $\kappa$.
These intensities are represented by distinct curves: $\kappa=1/2$
(black dotted line), $\kappa=1/\sqrt{6}$ (blue dashed line), and
$\kappa=1/(2\sqrt{2})$ (red solid line).
As expressed in Eq.\eqref{eq:psitauq}, at $t=\tau_{q}$, we have the
death of entanglement between the two optical and the mechanical modes.

Unfortunately, the von Neumann and linear entropy applicability as an
entanglement quantifier for mixed quantum states is limited.
Its limitation is its inability to discern between classical and quantum
correlations, meaning a nonnull value for the von Neumann entropy may
occur for separable mixed states.
When considering only the state of optical modes A and B, it is
represented by the reduced density operator
$\rho_{\text{AB}}(t)=\Tr_{\text{C}}\vert\psi(t)\rangle\langle
\psi(t)\vert$, obtained by tracing over all the degree of freedom of the
mirror, which typically results in a mixed state and prevents us of
using the von Neumann or the linear entropies.
However, defining an entanglement quantifier for mixed states in a
continuous-variables system can be challenging.
In those cases, employing inseparability criteria can be helpful for
determining whether the state of the system is or is not entangled
according to the adopted criterion.
Examples of inseparability criteria for continuous-variables systems are
presented in the Refs. \cite{duan2000,hillery2010}.

\begin{figure}[t]
\centering
\includegraphics[width=0.75\linewidth]{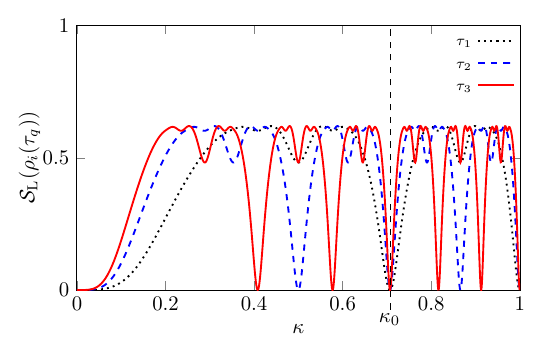}
\caption{
  Linear entropy of state of the $i$-th partition($i=\text{A},\text{B}$)
  as a function of the coupling constant $\kappa$ at different
  dimensionless times $t=\tau_{q}=2q\pi$ considering $\alpha=\beta=1$.
  In this case, the linear entropy quantifies the degree of entanglement
  between the two optical modes at times $\tau_{1}=2\pi$ (black dotted
  line), $\tau_{2}=4\pi$ (blue dashed line), and  $\tau_{3}=6\pi$ (red
  solid line).
  An interesting behavior occurs, particularly when
  $\kappa=\kappa_{0}=1/\sqrt{2}$ (vertical dashed line) and $\kappa=1$,
  for those nonnull coupling values, the tripartite system is fully
  separable at any time $t=\tau_{q}$ (see Eqs.\eqref{sepState1} and
  \eqref{sepState2}).
}
\label{fig:entropyAB_CS}
\end{figure}

Then, we can focus our analysis by restricting our state to the specific
instant of time $t=\tau_q$, allowing us to consider the system in a pure
state for the coupled optical modes described in Eq. \eqref{ABstate_q}.
At this time, the state of the system is represented by the density
operator $\rho_{\text{AB}}(\tau_{q})
= \vert \chi (\tau_{q})\rangle\langle\chi (\tau_{q})\vert$
(for simplicity, we have dropped the AB index from bracket notation).
Therefore, considering $\rho_{\text{AB}}(\tau_{q})$, we can write an
analytical expression for the linear entropy of modes A and B as follows
\begin{align}
  \label{SLa_coherent}
  \mathcal{S}_{\text{L}}(\rho_{i}(\tau_{q}))
  =1-
  \sum_{k,l,m,n=0}^{\infty}|c_{kl}|^{2}|c_{mn}|^{2}
  \cos{[2k_{a}k_{b}(k-m)(l-n)\tau_{q}]},
\end{align}
with $i = \text{A}, \text{B}$ indicating one of the optical systems.
To analyze this expression, we assume
$k_{\text{a}}\approx k_{\text{b}}=\kappa$ and nonnull values for
coherent states parameters $\alpha$ and $\beta$.
In the range of the coupling intensity $\kappa\in[0,1]$, we observe
separability for a nonnull coupling intensity when $\kappa=1/\sqrt{2}$,
where the state becomes
\begin{align}
\label{sepState1}
  \vert \chi(\tau_{q})\rangle_{\text{AB}} = {}
  &
    \frac{1+e^{\tau_{q}/2}}{2}
    |\alpha(\tau_{q})\rangle \otimes|\beta(\tau_{q})\rangle
    +
    \frac{1-e^{\tau_{q}/2}}{2}
    |-\alpha(\tau_{q})\rangle
    \otimes
    |-\beta(\tau_{q})\rangle,
\end{align}
and when $\kappa = 1$, the state is given by
\begin{align}
\label{sepState2}
  \vert \chi(\tau_{q})\rangle_{\text{AB}} = {}
  &
    |\alpha(\tau_{q})\rangle \otimes|\beta(\tau_{q})\rangle,
\end{align}
at any time $t=\tau_{q}$, as illustrated in Fig. \ref{fig:entropyAB_CS} for the case $\alpha=\beta=1$.
When $t\in(\tau_{q},\tau_{q+1})$ the continuous-variable mixed state $\rho_{\text{AB}}$ may be entangled.
Furthermore, from Fig. \ref{fig:entropyAB_CS}, we also observe, at least
during those times ($t=\tau_{q}$), that stronger optomechanical coupling
intensities do not necessarily imply higher entanglement between the
states of the optical fields.
Cases in which the entanglement has a periodic behavior may be
interesting to quantum information protocols for the application of
quantum gates on optical qubits, as suggested in Ref. \cite{asjad2015},
wherein the authors implement a deterministic quantum phase gate between
optical qubits associated with the two intracavity modes.

\section{Dissipative Dynamics}
\label{sec:nonunitary}

In this section, we investigated the dynamics when the optical
loss is negligible concerning the damping of the mirror.
It is possible, in principle, by choosing sufficiently reflective
mirrors; in that case, the mechanical damping rate would be some orders
of magnitude bigger than the damping rate due to light leakage.
Here, we consider the mirror weakly coupled to a thermal bath composite
by an infinite number of harmonic oscillators under the Born-Markov
approximation and consider the environment at zero temperature (only
dissipation occurs).
These assumptions lead us to the Gorini-Kossakowski-Sudarshan-Lindblad
(GKSL) master equation \cite{GKS1976,lindblad1976} to describe the
time-evolution of the density operator of the system, which can be
written as
\begin{align}
  \label{ME}
  \frac{d\rho^{\gamma}(t)}{dt}
  &
    =-i\left[\frac{H}{\hbar\omega_{\text{m}}},\rho^{\gamma}(t)\right]
    +\frac{\gamma}{2}
    \left[
    2c\rho^{\gamma}(t)c^{\dagger}-c^{\dagger}c\rho^{\gamma}(t)
    -\rho^{\gamma}(t)c^{\dagger}c
    \right],
\end{align}
where we consider the dimensionless time $\omega_{\text{m}}t \to t$, and
the dimensionless decoherence parameter
$\gamma = \gamma_{\text{diss}}/\omega_{\text{m}}$
represents the rate at which the system dissipates energy.
For more details about the quantum GKSL master equation, see Ref.
\cite{manzano2020}.

To solve the master equation in \eqref{ME}, we assume that the mirror
starts in a coherent state.
Consequently, we can apply the following ansatz \cite{sougato1997}
\begin{equation}
  \label{rho_sol}
  \rho^{\gamma}(t)=\sum_{k,l,m,n}\rho_{km,ln}^{\gamma}(t)\vert
  k\rangle\langle m\vert
  \otimes\vert l\rangle\langle n\vert\otimes
  \vert\phi_{kl}^{\gamma}(t)\rangle\langle\phi_{mn}^{\gamma}(t)\vert,
\end{equation}
into Eq. \eqref{ME} to solve the master equation.
After straightforward algebraic manipulation and  comparing term by
term, it provides the following differential equations
\begin{equation}
\label{Eq_phi}
\dot{\phi}_{kl}^{\gamma}+\left(i+\frac{\gamma}{2}\right)\phi_{kl}^{\gamma}
=i\kappa_{kl},
\end{equation}
and
\begin{align}
\label{rho_eq}
\frac{\dot{\rho}_{km,ln}^{\gamma}}{\rho_{km,ln}^{\gamma}} = & {}
-i
\Big\{
r_{a}(k-m)+r_{b}(l-n)
-
\kappa_{kl}\Re(\phi_{kl}^{\gamma})
+
\kappa_{mn}\Re(\phi_{mn}^{\gamma})
- \gamma\Im(\phi_{kl}^{\gamma}\phi_{mn}^{\gamma\ast})
\Big\}
\nonumber\\
&
-\frac{\gamma}{2}|\phi_{kl}^{\gamma} - \phi_{mn}^{\gamma}|^{2} .
\end{align}
The imaginary component accounts for the oscillations, while the real
component describes the exponential decay in the density operator
elements caused by the thermal bath.

Considering the mirror starts in a coherent state such that
$\phi_{kl}^{\gamma}(0)=\phi$, Eq. \eqref{Eq_phi} can be solved yielding
\begin{equation}
  \label{phi_g}
  \phi_{kl}^{\gamma}(t)=\phi e^{-(i+\gamma/2)t} + \kappa_{kl}\eta^{\gamma}(t),
\end{equation}
where $\eta^{\gamma}(t) \equiv [i/(i+\gamma/2)][1-e^{-(i+\gamma/2)t}]$.
Now we can integrate the Eq. \eqref{rho_eq} to obtain the density
operator coefficients
\begin{align}
\label{rho_elements}
    \rho_{km,ln}^{\gamma}(t) = {}
    &
    \rho_{km,ln}^{\gamma}(0)e^{-it\left[r_{a}(k-m)+r_{b}(l-n)\right]}
    \nonumber\\
    &
    \times
    e^{i(\kappa_{kl}-\kappa_{mn})\zeta_{\phi}^{\gamma}(t)}
    e^{i(\kappa_{kl}^{2}-\kappa_{mn}^{2})\Re[\xi^{\gamma}(t)]}
    e^{-(\kappa_{kl}-\kappa_{mn})^{2}\Gamma^{\gamma}(t)},
\end{align}
in which we introduced
  \begin{align}
    \zeta_{\phi}^{\gamma}(t) = {}
    &
    |\phi|
    \Big\{
    t\cos\varphi
      -\Big(\cos{\varphi}+\frac{3\gamma}{2}\sin{\varphi}\Big)
      \Re[\xi^{\gamma}(t)]
    \nonumber\\
    &
      +\Big(\sin{\varphi}+\frac{\gamma}{2}\cos{\varphi}\Big)
      \Im[\xi^{\gamma}(t)]
      +2\left(\sin{\varphi}-\frac{\gamma}{2}\cos{\varphi}\right)
      \Gamma^{\gamma}(t)
    \Big\}
    ,
  \end{align}
with $\varphi$ being the phase of $\phi$ in the polar form
$\phi=|\phi|e^{i\varphi}$.
Additionally, we defined the functions
$\xi^{\gamma}(t) = \int_{0}^{t}d\tau\eta^{\gamma}(\tau) =
\Re[\xi^{\gamma}(t)]+i\Im[\xi^{\gamma}(t)]$ and
$\Gamma^{\gamma}(t)=\frac{\gamma}{2}
\int_{0}^{t}d\tau|\eta^{\gamma}(\tau)|^2$
which have their explicit forms given by
\begin{subequations}
  \begin{align}
    \label{coh_term}
    \Re[\xi^{\gamma}(t)] = {}
    &
      \frac{1}{1+\gamma^{2}/4}
      \Bigg[
      t - \frac{1-\gamma^{2}/4}{1+\gamma^{2}/4}\mathfrak{s}_{\gamma}(t)
      -\frac{\gamma}{1+\gamma^{2}/4} (1-\mathfrak{c}_{\gamma}(t))
      \Bigg],
    \\
    \Im[\xi^{\gamma}(t)] = {}
    &
      \frac{1}{1+\gamma^{2}/4}
      \Bigg[
      \frac{\gamma t}{2} + \frac{1-\gamma^{2}/4}{1+\gamma^{2}/4}(1-\mathfrak{c}_{\gamma}(t))
      -\frac{\gamma}{1+\gamma^{2}/4}\mathfrak{s}_{\gamma}(t)
      \Bigg],
    \\
    \label{diss_term}
    \Gamma^{\gamma}(t) = {}
    &
      \frac{1}{1+\gamma^{2}/4}
      \Bigg[
      \frac{\gamma t}{2}+\frac{1-e^{-\gamma t}}{2}
      - \frac{\gamma}{1+\gamma^{2}/4} \mathfrak{s}_{\gamma}(t)
      -\frac{\gamma^{2}/2}{1+\gamma^{2}/4}(1-\mathfrak{c}_{\gamma}(t))
      \Bigg],
  \end{align}
\end{subequations}
where $\mathfrak{s}_{\gamma}(t)=e^{-\gamma t/2}\sin{t}$ and
$\mathfrak{c}_{\gamma}(t) = e^{-\gamma t/2}\cos t$.
It is important to point out that our method of solving the master
equation \eqref{ME} is exact, improving on the method used in
Ref. \cite{sougato1997}, wherein an approximate solution for $\phi=0$
neglecting terms proportional to $\mathcal{O}(\gamma)$ in
Eq. \eqref{coh_term} was obtained, i.e., $\Re[\xi(t)]\approx
t-\sin{t}$. Our solution includes any initial coherent state for the
mirror characterized by the parameter $\phi=|\phi|e^{i\varphi}$.

Throughout time evolution, the initial state $\rho^{\gamma}(0)$
converges towards a state of equilibrium with the thermal reservoir,
such that for a sufficiently long time ($t\to \infty$), the system
evolves to the steady-state
$\rho_{\infty}^{\gamma}\equiv\rho^{\gamma}(t\to\infty)$ given by
\begin{equation}
  \label{rho_equilibrium}
  \rho_{\infty}^{\gamma}
  =
  \sum_{k,l}\rho_{kk,ll}^{\gamma}(0)\vert k\rangle\langle k\vert
  \otimes
  \vert l\rangle\langle l\vert
  \otimes  \vert\kappa_{kl}\eta_{\infty}^{\gamma}\rangle
  \langle\kappa_{kl}\eta_{\infty}^{\gamma}\vert,
\end{equation}
with
$\eta_{\infty}^{\gamma}\equiv \eta^{\gamma}(t\to{\infty})=i/(i+\gamma/2)$.
The steady-state represents an equilibrium condition where the
properties of the system no longer change over time
\cite{SteadyStateGKSL}.
In other words, the flow of information or energy between the system and
its environment has balanced out.
This results from dissipative dynamics, where the system loses energy or
coherence due to environmental interaction, which leads to effects like
decoherence, where quantum superpositions are lost, and the system
behaves more classically.

\subsection{Vacuum one-photon superposition states}

Once we have derived the density operator governing the dissipative
dynamics of the optomechanical system, we can quantify the entanglement
between the optical partitions.
Specifically, we are interested in analyzing the scenario where the
fields are initially prepared in a separable non-Gaussian state,
comprising a superposition of vacuum and single-photon states, while the
mirror remains in a vacuum state.
This preparation holds significance due to the dynamics of the system
becoming confined to a two-dimensional space spanned by the vacuum and
one-photon states \cite{lombardi2002,brandao2020}.
Then, we consider the initial state as being
\begin{align}
  \vert\psi(0)\rangle =
  \vert + \rangle
  \otimes
  \vert + \rangle
  \otimes
  \vert 0 \rangle,
\end{align}
where we use the compact notation
$\vert+\rangle = (\vert 0 \rangle + \vert 1 \rangle)/\sqrt{2}$.
Note that the states of optical modes behave like a pair of qubits in
the way that the states $\vert0\rangle$ and $\vert1\rangle$ are the
eigenvectors of the Pauli matrix
$\sigma_z=\vert0\rangle\langle0\vert-\vert1\rangle\langle1\vert$ such
that $\sigma_z\vert0\rangle=+\vert0\rangle$ and
$\sigma_z\vert1\rangle=-\vert1\rangle$.
The mean-photon number in this state at each cavity is given by
$\langle a^{\dagger}a\rangle (0) = \langle b^{\dagger} b\rangle (0) = 1/2$,
which means that we have a single photon that can be found or not in one
of the cavities.
Moreover, the initial elements of the density operator are reduced to $\rho^{\gamma}_{km,ln}(0)=1/4$ for $k,m,l,n = 0,1$.

The interaction between the optical modes can be analyzed by eliminating
the degrees of freedom of the mirror, taking the trace over all of them
to obtain the reduced density operator
\begin{align}
\label{qubitRho}
  \rho^{\gamma}_{\text{AB}}(t) =
  \sum_{k,l,m,n=0}^{1}\rho_{km,ln}^{\gamma}(t)
  \langle\phi_{mn}^{\gamma}(t)\vert \phi_{kl}^{\gamma}(t)\rangle
  \vert k\rangle\langle m\vert\otimes\vert l\rangle\langle n\vert,
\end{align}
where
$\langle\phi_{mn}^{\gamma}(t)\vert \phi_{kl}^{\gamma}(t)\rangle =
e^{-|\phi_{kl}^{\gamma}(t)-\phi_{mn}^{\gamma}(t)|^2/2}
e^{-i\Im{[\phi_{kl}^{\gamma}(t)\phi_{mn}^{\gamma\ast}(t)]}}$.
Also, we can trace the degrees of freedom of optical modes to obtain the
state of the mirror
\begin{align}
\label{rho_mirror}
  \rho^{\gamma}_{\text{C}}(t)=
  \sum_{k,l=0}^{1}\rho_{kk,ll}^{\gamma}(t) \vert\phi_{kl}^{\gamma}(t)
  \rangle \langle\phi_{kl}^{\gamma}(t)\vert.
\end{align}
We observe that $\rho^{\gamma}_{\text{C}}(t)$ is represented by a convex
sum of coherent states, being diagonal in this basis and representing a
classical state in this sense.
\begin{figure}[t]
  \centering
  \includegraphics[width=0.75\linewidth]{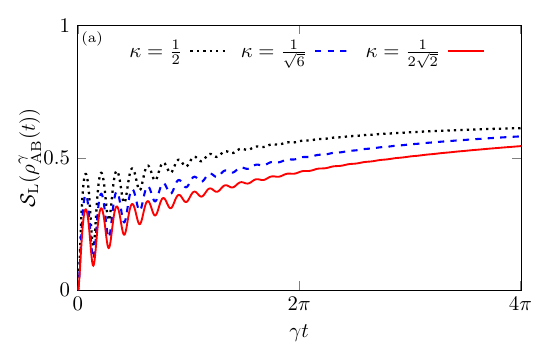}
  \includegraphics[width=0.75\linewidth]{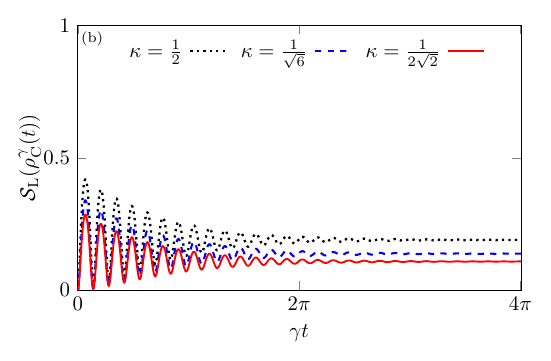}
  \caption{
    The linear entropies of partitions AB and C are plotted as functions
    of scaled time $\gamma t$, by setting the decay constant
    $\gamma=0.07$.
    Here, linear entropy only quantifies the degree of purity of the
    optical modes state in (a) and the mechanical state in (b) by
    considering different optomechanical coupling values
    $k_{\text{a}}\approx k_{\text{a}}=\kappa$ with the curves
    representing the cases $\kappa=1/2$ (black dotted line),
    $\kappa=1/\sqrt{6}$ (blue dashed line), and $\kappa=1/(2\sqrt{2})$
    (red solid line).
  }
  \label{fig:entropyQubits}
\end{figure}

Unlike the unitary case discussed before, now the system cannot be
separated as a direct product of the optical modes with the states of
the mirror at any time because of the effect of the environment that
couples the modes and the mirror at any instance.
Therefore, once the system evolves to a mixed state, the linear entropy
does not represent a suitable quantifier of entanglement, although it is
still a quantifier of the purity of the states.
In such cases, two separable systems that are not entangled with each
other can have nonzero entropy since entropy not only takes into account
the quantum entanglement but also the classical correlation
\cite{horodecki2009}.
Then, we compute the linear entropies for the optical modes state and
the mirror state, which are respectively expressed as follows
\begin{align}
  S_{\text{L}}(\rho^{\gamma}_{\text{AB}}(t))
  = 1- \frac{1}{16}\sum_{k,l,m,n=0}^{1}
  e^{-(\kappa_{kl}-\kappa_{mn})^2[|\eta^{\gamma}(t)|^{2}
  + 2\Gamma^{\gamma}(t)]},
\end{align}
and
\begin{equation}
  S_{\text{L}}(\rho^{\gamma}_{\text{C}}(t))
  = 1 - \frac{1}{16}\sum_{k,l,m,n=0}^{1}
  e^{-(\kappa_{kl}-\kappa_{mn})^2|\eta^{\gamma}(t)|^{2}}.
\end{equation}
Note that the expression of the AB system is independent of the phase of
the function
$\langle\phi_{mn}^{\gamma}(t)\vert \phi_{kl}^{\gamma}(t)\rangle$
and the coherent state parameter $\phi$, and therefore, the result is
the same obtained by the approximated method applied in
\cite{sougato1997}.
These expressions are plotted in Fig. \ref{fig:entropyQubits} in which
we set the decay constant $\gamma=0.07$, and analyze the dynamics of the
linear entropies as functions of the re-scaled time $\gamma t$ for
different coupling intensities under the assumption
$k_{\text{a}}\approx k_{\text{b}} = \kappa$.
We verify that the more the coupling between the optical modes and
mirror increases, the more the degree of purity decreases, which is
expected once the mirror is connected to the environment, leading to
coherence loss.

Furthermore, there is a distinct advantage of employing discrete states
such as qubits in studying entanglement.
By doing so, we can avoid the problems associated with employing linear
entropy as an entanglement quantifier and instead utilize concurrence
\cite{wootters1998} as an accurate measure of entanglement, following
the work done in \cite{brandao2020}, where the undamped case was
considered.
The concurrence for a state of two-qubits $\rho$ is defined as
\begin{equation}
  \label{concurrence}
  \mathcal{C}(\rho)
  = \text{max}[0,\lambda_1 - \lambda_2 - \lambda_3 -\lambda_4],
\end{equation}
where $\lambda_i$ are square roots of the eigenvalues, in decreasing
order, of the non-Hermitian matrix $R= \rho\tilde{\rho}$ with
\begin{equation}
  \tilde{\rho} = (\sigma_y \otimes \sigma_y)\rho^{\ast}
  (\sigma_y \otimes \sigma_y),
\end{equation}
being the spin-flipped density matrix with
$\sigma_{y}=i(\vert1\rangle\langle0\vert-\vert0\rangle\langle1\vert)$,
where $\rho^{\ast}$ is the complex conjugate of a given state $\rho$.

\begin{figure}[t]
  \centering
  \includegraphics[width = 0.75\linewidth]{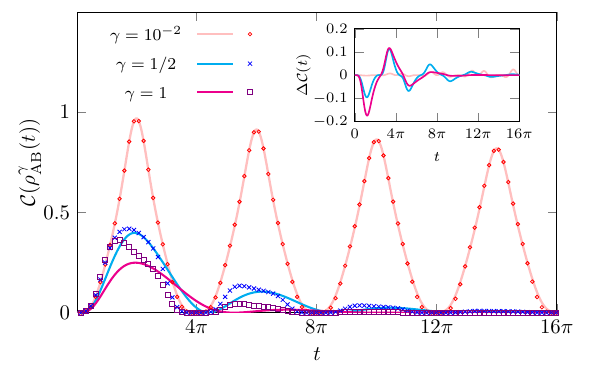}
  \caption{
    Concurrence of $\rho_{\text{AB}}^{\gamma}(t)$ as a function of
    dimensionless time $t$ for $\kappa=1/2$.
    The solid lines represent the concurrence calculated from the exact
    solution obtained in our work, while the markers represent the
    concurrence obtained from the approximated solution obtained in
    Ref. \cite{sougato1997}.
    The difference between them, namely $\Delta\mathcal{C}(t)$ is
    depicted in the inset.
    The behavior of concurrence is analyzed for different values of
    decay parameter $\gamma$.
    For $\gamma =10^{-2}$, the approximation (red circles) matches the
    exact (pink line) case.
    For $\gamma=\kappa$, the approximation (blue crosses) deviates from
    the exact solution (cyan line).
    For $\gamma=1$, the approximation (violet squares) also fails to
    represent the exact result (magenta line).
  }
  \label{fig:concurrenceQubits}
\end{figure}

To obtain the concurrence between the optical field states, we
explicitly write the reduced density operator $\rho_{\text{AB}}(t)$ of
Eq. \eqref{qubitRho} in the matrix form
\begin{equation}
\!\!\! \rho^{\gamma}_{\text{AB}}(t) \! = \!
\frac{1}{4}\!\!
  \begin{bmatrix}
    1      &    b(t)    &    a(t)    &    a(t)b(t)e^{\theta(t)}
    \\
    b^{\ast}(t)    &    1       &    a(t)b^{\ast}(t)e^{-\Re[\theta(t)]}    &    a(t)e^{i\Im[\theta(t)]}
    \\
    a^{\ast}(t)    &     a^{\ast}(t)b(t)e^{-\Re[\theta(t)]}    &    1    &    b(t)e^{i\Im[\theta(t)]}
    \\
    a^{\ast}(t)b^{\ast}(t)e^{\theta^{\ast}(t)}    &    a^{\ast}(t)e^{-i\Im[\theta(t)]}    &    b^{\ast}(t)e^{-i\Im[\theta(t)]}    &    1
  \end{bmatrix}\!\!,
\end{equation}
where we define the functions
\begin{subequations}
\begin{align}
a(t)
= {}
&
e^{ir_{\text{a}}t}
e^{-ik_{\text{a}}[\zeta_{\phi}^{\gamma}(t)-\mu_{\phi}^{\gamma}(t)]}
e^{-ik_{\text{a}}^{2}\Re{[\xi^{\gamma}(t)]}}
e^{-k_{\text{a}}^{2}[|\eta^{\gamma}(t)|^{2}+2\Gamma^{\gamma}(t)]/2},
\\
b(t)
= {}
&
e^{ir_{\text{b}}t}
e^{ik_{\text{b}}[\zeta_{\phi}^{\gamma}(t)-\mu_{\phi}^{\gamma}(t)]}
e^{-ik_{\text{b}}^{2}\Re{[\xi^{\gamma}(t)]}}
e^{-k_{\text{b}}^{2}[|\eta^{\gamma}(t)|^{2}+2\Gamma^{\gamma}(t)]/2},
\\
\theta(t) = {}
&
k_{\text{a}}k_{\text{b}}
  \left\{|\eta^{\gamma}(t)|^{2}+2\Gamma^{\gamma}(t)
  + 2i\Re[\xi^{\gamma}(t)]\right\},
\end{align}
\end{subequations}
with
\begin{align}
\mu_{\phi}^{\gamma}(t) = |\phi|
\left[
\cos{\varphi}\Im[\eta^{\gamma}(t)]-\sin{\varphi}\Re[\eta^{\gamma}(t)]
+\left(\sin{\varphi}-\frac{\gamma}{2}\cos{\varphi}\right)|\eta^{\gamma}(t)|^{2}
\right].
\end{align}
Hence, we can quantify the entanglement by employing the concurrence
definition in Eq. \eqref{concurrence}.
A numerical analysis of the behavior of concurrence
$\mathcal{C}(\rho_{\text{AB}}^{\gamma})$ against the dimensionless time
$t$ shows that the concurrence is independent of the parameters
$r_{\text{a}}$, $r_{\text{b}}$, and $\phi$.
This observation is consistent with the property that concurrence is
invariant under local unitaries \cite{wootters2001}.
As represented in Fig. \ref{fig:concurrenceQubits}, our analysis
considers the exact (solid lines) density matrix obtained in this work
and the approximated (marked lines) one, which is obtained employing the
method reported in Ref. \cite{sougato1997}.
Furthermore, we defined the difference between the exact and
approximated concurrences, namely $\Delta\mathcal{C}(t)$, represented in
the inset at Fig. \ref{fig:concurrenceQubits}.
This difference may be of interest when characterizing the parameters
and time intervals and determining whether the approximation is
reasonable.
Fixing an optomechanical coupling $k_{\text{a}}\approx
k_{\text{b}}=\kappa=1/2$, we compare the concurrences for different
values of the decay constant $\gamma$.
As expected, the discrepancies between the concurrences generated by the
exact and approximated density operators become more pronounced with the
increase of $\gamma$.
For small values of $\gamma$, for instance, $\gamma=10^{-2}$,  we
observe that the approximation (red circles) closely matches the exact
solution (pink line) with $\Delta\mathcal{C}(t)\approx0$.
However, when $\gamma$ is of the same order of magnitude or larger than
the optomechanical coupling intensity, we notice a more significant
discrepancies between the approximate and exact concurrences.
For $\gamma=\kappa$, the approximation (blue crosses) deviates
noticeably from the exact solution (cyan line).
Similarly, for $\gamma=1$, the approximation (violet squares) also fails
to accurately represent the exact solution (magenta line).
In those cases, we observe a difference between exact and approximated
concurrence $\Delta\mathcal{C}(t)$, exceeding $10\%$ in magnitude up to
$t=4\pi$.
Beyond that, the difference decreases as both quantities approach zero
for larger times.
In the scenario where the mechanical loss is introduced ($\gamma \neq
0$), the birth and death of entanglement persist, albeit gradually
attenuated over time up to when the system reaches the steady state
$\rho_{\infty}^{\gamma}$ (see Eq. \eqref{rho_equilibrium}).

Regarding an experimental implementation of this system, there are no
actual experiments exploiting the properties of the mirror-in-the-middle optomechanical system discussed above.
Nevertheless, detailed experimental proposes were developed in recent
years such as the ones by Brandão  et al. \cite{brandao2020,brandao2021}
and Kanari-Naish et al. \cite{kanari2022}.
Those suggest interest in implementing this kind of system in the lab
and the proposals utilizing this model are rich and varied, ranging from levitated nanospheres, to ultracold atomic ensembles and a Mach–Zehnder
interferometer containing two optomechanical cavities.

\section{Conclusion}
\label{sec:conclusion}

In this work, we studied the mirror-in-the-middle optomechanical system
featuring mechanical loss, where the movable mirror operates within the
framework of quantum mechanics.
In the absence of losses, we witness the transition of an initially
separable state composed of the product of coherent states into an
entangled one, revealing the emergence and decay of entanglement
for continuous-variable states through the analysis of linear entropy.
This evolution highlights the inherently nonclassical behavior of the
mechanical oscillator within this context.
We explicitly demonstrate the generation of a two-mode multi-component
Schrödinger-cat state depending on the optomechanical coupling at
dimensionless times $t=\tau_{q}=2q\pi$ with $q\in\mathbb{N}$.
During these instances, the global state of the optical fields remains
disentangled from the mirror state.
However, the optical fields may be entangled, allowing us to quantify
this entanglement as a function of the optomechanical coupling.
An intriguing observation emerges in these instances: the optical fields
exhibit separability at specific nonnull coupling intensities, namely
$\kappa = 1/\sqrt{2}$ and $\kappa = 1$.

When the mechanical loss is considered, the GKSL master equation is
exactly solved by applying the ansatz described in Eq. \eqref{rho_sol}.
This ansatz is also considered in Ref. \cite{sougato1997}.
However, the authors utilized alternating unitary and nonunitary
evolutions in short intervals to tackle the master equation, whereas our
approach directly engages with the differential equations.
As a result, we improved their solution, showing that while the damping
term remains consistent in both methodologies, our approach highlights
the influence of the reservoir on the coherent term.
Additionally, we utilize the exact solution to evaluate entanglement
between the optical fields when they are prepared in vacuum one-photon
superposition states, where the dynamics of each field are constrained
to the two-dimensional subspace spanned by
$\{\vert0\rangle,\vert1\rangle\}$, and concurrence may be employed as a
quantifier of entanglement.
Then, we verified how entanglement in states of the optical fields is
attenuated when the mechanical loss is considered.
Furthermore, we compare the concurrence obtained from our exact density
matrix with the approximation given in Ref. \cite{sougato1997}, and we
certify that both results match only for small decay parameter $\gamma$.

As a future task, an analysis of the possibility of generating
tripartite entanglement among the partitions in this configuration may
be done \cite{aoki2003}.
Finally, we believe our findings may complement previous analyses in an
exact description of optomechanical dynamics by including mechanical
loss.
It offers a natural step forward in the results reported in the
literature, e.g., Ref. \cite{brandao2020}, where the authors explore
nonclassical features on optomechanical systems in the absence of
losses, and propose an experiment employing ultracold atomic ensembles.
Furthermore, from the perspective of the foundations of quantum
mechanics, it would be valuable to investigate other nonclassical
features of optomechanical systems, such as Bell-nonlocality, shown to
be possible in optomechanical systems and experimentally demonstrated in
\cite{marinkoviv2018}.

\section*{Acknowledgments}
DC would like to thank Prof. Bárbara Amaral and the Programa de Estímulo
à Supervisão de Pós-Doutorandos por Jovens Pesquisadores da Pró-reitoria
de Pesquisa e Inovação for providing the postdoctoral fellowship through
the financial support of Instituto Serrapilheira, Chamada 2020, and
Fundação de Auxílio à Pesquisa de São Paulo (FAPESP) - Jovem
Pesquisador, grant number 2020/06454-7.
This work was supported by the Brazilian agencies Conselho Nacional de
Desenvolvimento Cient\'ifico e Te\-cnol\'ogico (CNPq) and Instituto
Nacional de Ci\^{e}ncia e Tecnologia de Informa\c{c}\~{a}o Qu\^{a}ntica
(CNPq, INCT-IQ Grant No. 465469/2014-0).
This work was partially supported by Co\-or\-dena\c{c}\~{a}o de
Aperfei\c{c}oamento de Pessoal de N\'{i}vel Superior (CAPES, Finance
Code 001).
FMA acknowledges financial support by CNPq Grant No. 313124/2023-0.

\section*{Data Availability Statement}
No Data is associated with the manuscript.

\section*{References}

\bibliographystyle{iopart-num}
\providecommand{\newblock}{}

\end{document}